\documentclass[traditabstract]{aa}

\usepackage{txfonts}
\usepackage{t1enc}
\usepackage{longtable}
\usepackage{natbib}
\usepackage{array}
\usepackage{graphicx}
\usepackage{verbatim}
\usepackage{float}

\begin{document}

\title{A catalogue of Paschen-line profiles in standard stars
  \thanks{Figures 7-96 are only available in electronic form via
	http://www.edpsciences.org}
 }

\author{Wenjin Huang,\inst{1,2}\thanks{Current email and mailing address: wenjin.huang@asml.com, Brion Technologies Inc., 4211 Burton Dr., Santa Clara, CA 95054}
George Wallerstein,\inst{1,2}
\and Myra Stone\inst{1}
}

\institute{Department of Astronomy\\
University of Washington, Box 351580, Seattle, WA 98195-1580;\\
\email{hwenjin@astro.washington.edu}, \email{wall@astro.washington.edu}
\and
Visiting Astronomer, Dominion Astrophysical Observatory, Herzberg Institute for Astrophysics, Victoria, B.C., Canada}

\date{Received  /
	Accepted }


\abstract{
We have assembled an atlas of line profiles of the Paschen Delta (P$\delta$) line at 10\,049 $\AA$ for the use of stellar modelling. For a few stars we have substituted the Paschen Gamma (P$\gamma$) line at 10\,938 $\AA$ because the P$\delta$ line blends with other features. Most of the targets are standard stars of spectral types from B to M. A few metal-poor stars have been included. For many of the stars we have also observed the Hydrogen Alpha (H$\alpha$) line so as to compare the profiles of lines originating from the meta-stable $n=2$ level with lines originating from the $n=3$ level. The greatest difference in line profile is found for high luminosity and cool stars where the departures from LTE in the population of the $n=2$ level is expected to be the greatest.
\\
For a few stars, sample line profiles have been calculated in the LTE approximation to demonstrate the usefulness of the tabulated and displayed catalogue.}


\date{Recieved / Accepted }

\titlerunning{A catalogue of Paschen-line profiles}
\authorrunning{Huang, Wallerstein, \& Stone}
\maketitle

\setcounter{footnote}{0}
\section{Introduction}                              
The theory of model stellar atmospheres dates from the plane-parallel, single-layer models of \cite{sch1902, sch1905} and Schwarzschild (1906). In 1939 two very important papers on stellar atmospheres were published by Spitzer and by Struve, Wurm, and Henyey. In Spitzer's (1939) paper on the atmospheres of M supergiants he noted that an excitation temperature of 17\,000K was necessary to explain the strength of the H$\alpha$ absorption line in $\alpha$Ori even though the effective temperature of the star was about 3\,400K. In Fig.\ref{rev_HD39801_color}, we show our recent spectrum of $\alpha$Ori with a resolving power of about 30\,000. The observed H$\alpha$ line (Fig.\ref{rev_HD39801_color}a) is prominent and does not appear in the spectrum as modelled with a standard atmosphere in local thermodynamic equilibrium (LTE). $\alpha$Ori's P$\delta$ line (Fig.\ref{rev_HD39801_color}b) is dominated by a blend of neutral lines, and its P$\gamma$ line (Fig.\ref{rev_HD39801_color}c) is obscured by a molecular band. \citet{str39} called attention to the important fact that the 2s level from which the Balmer lines arise is meta-stable with a very small probability for the 2s$\rightarrow$1s transition. This may result in an over-population of the 2s level as compared with what is expected by the Boltzmann distribution in LTE. We quote them as follows, "It is tempting to explain the great strength of the Balmer absorption lines in supergiants of late type as a consequence of the metastability of the 2s level. However, it is premature to discuss this matter further until accurate measurements of the Paschen lines have been made."

Most early work on stellar atmospheres was based on the assumption of LTE in which the relative populations of the atomic levels are controlled by the local temperature and the assumption that the radiation field is also dependent on the local temperature \citep{uns55}. In succeeding years the degree of realism of model atmospheres has steadily increased, first, with sophistically calculated gradients of various physical quantities  \citep{mih73}, through models with extended atmospheres, to the most recent models with surface fluctuations of all relevant quantities. The most recent models include surface variations associated with large scale convection and granulation as well as departures from LTE in their excitation and ionization equilibria \citep{asp05}.

As noted by \citet{str39}, in real stars there is a special problem with the assumption of LTE because the radiative transition from the 2s level of hydrogen to the 1s ground state is forbidden. This means that the population of the 2s level must be controlled by radiative and collisional transitions with other excited levels, but transitions between the 2s and 1s levels are strongly inhibited. At high densities collisions are effective so that the population ratios among the $n>2$ levels approach equilibrium but the relative populations between $n=1$ and $n=2$ deviate from LTE. The populations of all levels are influenced by recombination and ionization, especially in the hotter stars. There is no question as to the presence of non-LTE (NLTE) effects; the only question is one of degree. The Paschen lines which originate from the $n=3$ level are expected to be closer to LTE than are the Balmer lines. To investigate the importance of these effects it is helpful to observe the Paschen (and higher) lines and to compare their line profiles with the predictions of atmospheric models. An additional consideration is the increased Stark effect for the Paschen lines as compared to the Balmer lines because Stark broadening increases with higher quantum numbers. Hence, the Stark broadening for the Paschen lines is expected to dominate over other broadening mechanisms.


\section{Observations}                       

With these factors in mind we have been observing a wide variety of stars at both H$\alpha$ and 10\,049 $\AA$ P$\delta$. The data have been obtained almost entirely with the coud\'{e} spectrograph of the 1.2-m telescope at the Dominion Astrophysical Observatory (DAO) of the Hertzberg Institute of Astrophysics at Victoria British Columbia. Their 1.2-m telescope with a single order coud\'{e} spectrograph is the ideal instrument for a spectroscopic survey of bright stars in a limited wavelength region since it does not suffer from the rapidly varying continuum in each short order of an echelle system. The 96-inch camera provides a resolving power of about 30\,000 over 250 $\AA$ at each setting of the first order of the grating. The detector is a SITE 4 CCD with 24 micron pixels. There is a low level of fringing in the near infrared that can be removed by flat-fielding to a remaining noise level of about 1\%. Since the stars are bright, the remaining source of noise is proportional to the square-root of the number of electrons per pixel and to absorption by the earth's atmosphere. The latter can be largely eliminated by division of a rapidly rotating bright hot star, though caution must be used since almost all such stars have their own hydrogen lines. For broad lines whose H-line profiles are defined by many pixels the noise level is unimportant. Wavelength calibration in the near infrared can be difficult due to the paucity of emission lines in most comparison arcs. This was readily overcome by observing the ThAr spectrum in the second order and simply multipying the wavelengths so calibrated by exactly 2.

For a few stars of types K and M there is an FeI line that blends with P$\delta$ and the latter is quite weak. Since these stars are very bright in the near IR it was possible to observe the brighter ones with the SITE 4 at the P$\gamma$ line at 10\,938 $\AA$. The 10\,830 $\AA$ line of HeI is present in some of those spectra.

\addtocounter{table}{1}

To overcome the problem of the FeI blend with the 10\,049 $\AA$ line in the cooler stars we included some metal-poor stars. Most of those stars are too faint for the 1.2-m telescope so they were also observed with the Apache Point Observatory 3.5-m echelle. In Table \ref{t1} we list the stars for which useful spectra were obtained at the DAO.  We list the metal-poor stars observed at the Apache Point Observatory in Table \ref{t2}. In both tables the columns are self-explanatory. The $M_{v}$ values are derived from the parallaxes measured by Hipparcos \citep{van07} if the measured parallax is at least 3 times its quoted probable error. Actually, most of the parallaxes are at least 5 times their quoted error. An asterisk is shown for stars whose parallax lies between 3.0 and 5.0 times its uncertainty. No $M_{v}$ value is shown for stars with parallaxes of greater uncertainty.


\section{The line profiles}      		 

The primary purpose of this paper is to make available accurate line profiles of selected Paschen lines so they may be used in connection with modelling stellar atmospheres. In addition, the Paschen lines may be used for calibrating the temperature scale and analysing the chemical composition of stellar atmospheres where minimizing the influence of NLTE is helpful. In a series of tables we present the line profiles as observed. Most of the stars are represented by the P$\delta$ line at 10\,049 $\AA$. In a few stars, especially stars of type K and M, the FeI line blends with P$\delta$. As mentioned above, we have observed the P$\gamma$ line at 10\,938 $\AA$ whenever possible in addition to including the metal-poor stars with weakened FeI lines. For many of the stars we also obtained H$\alpha$ profiles which we show for comparison with the Paschen line profiles. In Table \ref{t3} we list the wavelength, excitation of the lower level, and the transition probability of the three lines in this study. Our complete digital spectra have been deposited with the Centre de Donn\'{e}es astronomiques de Strasbourg for access to all astronomers. To provide a complete atlas of our observations we have included in the electronic edition all of our line profiles as figures. In the printed edition we illustrate a variety of line profiles for a wide range of spectral types.

\section{LTE profile comparisons} 	

We also carried out calculations of H$\alpha$ and P$\delta$ profiles for some stars using the Kurucz models in LTE \citep{kur93,sbo04,sbo05}, the ATLAS atmospheric model production code, and the SYNTHE spectral synthesis code \citep{kur93,sbo04}. The Stark broadened profiles in the SYNTHE code are "based on the quasi-static Griem theory with parameters adjusted in such a way that profiles from the Griem theory fit the [\citet{vidal73}] profiles of the first members of the Lyman and Balmer series" \citep{cowley02}. Stellar parameters for profile modelling (see Table \ref{t4}) were obtained from references in the online astronomical database SIMBAD with favorable weight given to the data of \citet{fulbright00}. Further details about the tools for atmosphere model preparation and line synthesis can be found online at the ATLAS, WIDTH and SYNTHE GNU Linux port\footnote{http://atmos.obspm.fr/} \emph{Home} page and under the \emph{Download} page.

The figures in the electronic edition show the synthetic LTE profiles plotted in red and the observed spectra in black. In the printed edition, the synthesized LTE profiles are plotted with dashed lines, while the observed spectra are in solid lines.

\subsection{Comments on selected stars shown in the printed version}		
\subsubsection{Spectral sequence at high luminosity with LTE profiles}		

HD37128, B0 Iab: H$\alpha$ (Fig.\ref{early_LumI_color}a) shows a typical P Cygni profile while P$\delta$ (Fig.\ref{early_LumI_color}g) is symmetric. 

HD46300 (13 Mon) A0 Ib: For 13 Mon there is a small difference in the profile of H$\alpha$ (Fig.\ref{early_LumI_color}b) while the Paschen line (Fig.\ref{early_LumI_color}h) fits the model almost exactly.

HD182835, F2 Iab: As for HD46300, there is a small difference between the model and the observations in the line wings (Fig.\ref{early_LumI_color}c). The Paschen line fits almost exactly (Fig.\ref{early_LumI_color}i).

HD209750, G2 Ib: The observed H$\alpha$ line (Fig.\ref{early_LumI_color}d) appears to be wider than the model, while the model and observation of the P$\delta$ line (Fig.\ref{early_LumI_color}j) fit very well.

HD206778, K2 Ib \& HD42543, M0 Iab: The trend as the spectral type becomes later shows the observed H$\alpha$ line (Fig.\ref{early_LumI_color}e \& Fig.\ref{early_LumI_color}f) becoming steadily stronger as compared to the model. 


\subsubsection{Sequence at high luminosity at mid-spectral types}		

HD34085 (Rigel) B8 Iab: This standard B supergiant shows a self-absorbed H$\alpha$ emission line (Fig.\ref{late_LumI_color}a) and a symmetric P$\delta$ line (Fig.\ref{late_LumI_color}g).

HD17378, A5 Ia: This supergiant member of the h and $\chi$ Per association is a little cooler than $\alpha$Cyg and shows an almost symmetric H$\alpha$ line (Fig.\ref{late_LumI_color}b) at the time of observation with little evidence for either a wind or active chromosphere. 

HD20902 ($\alpha$Per) F5 Iab: The fit for H$\alpha$ (Fig.\ref{late_LumI_color}c) is very good and for P$\delta$ (Fig.\ref{late_LumI_color}i) is almost perfect. 

HD77912, G8 Ib-II: The strong absorption at H$\alpha$ (Fig.\ref{late_LumI_color}d) contrasts with the weaker line at P$\delta$ (Fig.\ref{late_LumI_color}j).

HD44537, K5 Iab: The H$\alpha$ line (Fig.\ref{late_LumI_color}e) is strongly present and asymmetric. The P$\delta$ line (Fig.\ref{late_LumI_color}k) is dominated by a blend of neutral lines.

HD156014, M5 Ib-II: Despite the late spectral type, the observed H$\alpha$ profile (Fig.\ref{late_LumI_color}f) remains strong. 

\subsubsection{Spectral type K stars}		


HD122563, K0 IIp: This very metal-poor giant, whose [Fe/H] = -2.7, shows only small deviations from the model for its H$\alpha$ (Fig.\ref{SpT_K_color}b) and P$\delta$ (Fig.\ref{SpT_K_color}h) lines.

HD232078, K3 IIp: (Fig.\ref{SpT_K_color}c and Fig.\ref{SpT_K_color}i ) This red giant is in a markedly retrograde orbit and is similar to globular cluster red giants. Its $T_{\rm eff}$ of 4\,000K is one of the lowest values for a metal-poor field star. With [Fe/H] = -1.6, the metallic lines still obscure the P$\delta$ line.

HD124897 (Arcturus) K1.5 III: For this moderately metal-poor star, the observed H$\alpha$ line (Fig.\ref{SpT_K_color}d) is much stronger than the model profile but the P$\delta$ line (Fig.\ref{SpT_K_color}j) fits the model very well.

HD103095 (Groombridge 1830) K0 Vp: This well-known, metal-poor main sequence star, with a $T_{\rm eff}$ of 5\,000K, appears to have a slightly deeper core at H$\alpha$ (Fig.\ref{SpT_K_color}e) and shallower core at P$\delta$  (Fig.\ref{SpT_K_color}k) than predicted.

\subsubsection{P$\gamma$ profiles}		
HD172167 (Vega) A0 V: This fundamental standard for modelling stellar atmospheres shows a very good fit between the modelled and the observed profile for P$\gamma$ (Fig.\ref{pgamma_color}a).

HD31964, A8 Iab: The P$\gamma$ line is clearly shown (Fig.\ref{pgamma_color}b).

HD20902 ($\alpha$Per) F5 Iab: There is a great deal of blending at 10\,938 $\AA$, but the P$\gamma$ line is visible (Fig.\ref{pgamma_color}c).

HD206778, K1.5 III: The P$\gamma$ line is obscured by a molecular band (Fig.\ref{pgamma_color}d).

\subsection{Comments on selected stars shown in the electronic version only}		

Fig.15, HD886, B2 IV: This standard early B star shows symmetrical H$\alpha$ and P$\delta$ profiles.

Fig.41, HD197345 (Deneb) A2 Iae: This standard supergiant shows its well-known P Cygni structure at H$\alpha$ while its P$\delta$ profile fits the model exactly.

Fig.52, HD18391, G0 Ia: This G supergiant may be an outer member of the h and $\chi$ Per cluster. Its H lines appear to be symmetric.

Fig.69, HD6860, M0 III: The H$\alpha$ line is clearly present while the P$\delta$ line is dominated by the blend of neutral metals. There appears to be almost no feature at P$\gamma$. A weak molecular band may be interfering.
 
Fig.71, HD206936 ($\mu$Cep) M2 Ia: This supergiant may be even more luminous than $\alpha$Ori. Its H$\alpha$ line is similar to that of $\alpha$Ori. There is no evidence for the P$\gamma$ line though the molecular band seen in $\alpha$Ori is weaker in $\mu$Cep.

Fig.79, HD172380 (XY Lyr) M4 Iab: H$\alpha$ is clearly present while P$\gamma$ is not evident. The molecular band seen in $\alpha$Ori is not present.

Fig.80, HD197812 (U Del) M5 Ia: A weak molecular band appears in the star and in most stars of type M5 or later very close to the position of H$\alpha$. No line is visible at P$\gamma$. 


Fig.85, HD221170, G2 IV: H$\alpha$ is symmetrical and much stronger than P$\delta$.

Fig.94, HD25329, K1 V: This subdwarf is significantly metal-poor and is similar to HD103095. For both stars H$\alpha$ is much stronger than P$\delta$.

Fig.95, HD165195, K3 p: This star's metallicity is similar to giant branch stars in globular clusters and shows weak H$\alpha$ emission that is usually associated with mass loss.

\section{Discussion}   

Because the NLTE effects are much reduced in main-sequence stars, we expect that the synthesized LTE profiles should fit well with both H$\alpha$ and P$\delta$ observation. The result in Fig.\ref{rev_HD172167_color} (for Vega, A0 V) is a very good example of this case, the LTE model ( $T_{\rm eff}$ = 9\,540K, $log_{\rm g}$ = 4.0) fit both H$\alpha$ and P$\delta$ well. But for the A0 supergiant, 13 Mon (HD 46300, A0 Ib), an LTE model can not match them simultaneously. When the LTE model matches the observed P$\delta$ well (see Fig.\ref{early_LumI_color}h), it overestimates the strength of H$\alpha$ (see Fig.\ref{early_LumI_color}b). 
 
For super giants with lower temperature (but hotter than F5),  we continue to see a similar difference between H$\alpha$ and P$\delta$ fittings which, however, become smaller towards lower temperatures (see Fig.\ref{early_LumI_color}c and Fig.\ref{early_LumI_color}i). 
The difference disappears around the spectral type of F5 ($\alpha$Per, F5 Iab, in Fig.\ref{late_LumI_color}c and Fig.\ref{late_LumI_color}i, for example).

When the effective temperature drops to that of G type, the LTE models once again show their failure to reproduce the observed H$\alpha$ profiles. In contrast to the case of hot luminous stars (Fig.\ref{rev_HD172167_color}, Fig.\ref{early_LumI_color}b, and Fig.\ref{early_LumI_color}h), this time the LTE model profiles of H$\alpha$ are weaker than the observed line profiles (Fig.\ref{early_LumI_color}d).

In the low temperature regime of K and M type giants/supergiants, the H$\alpha$ profiles predicted by the LTE models become very weak while the observed profiles maintain their strength to appear almost the same as this of G stars. Arcturus, a moderately metal-poor star, has a much stronger H$\alpha$ line than modelled (Fig.\ref{SpT_K_color}d). The very metal-poor giant HD122563 also shows a weaker LTE-modelled H$\alpha$ profile (Fig.\ref{SpT_K_color}b), as does the metal-poor K0 dwarf Groombridge 1830 (Fig.\ref{SpT_K_color}e). In later type M giants, the spectrum is dominated by TiO bands and H$\alpha$ disappears. We should note that, for those cool luminous stars with detectable P$\delta$, the synthetic P$\delta$ LTE profiles still provide a good fit to the observed spectra (Fig.\ref{SpT_K_color}h and Fig.\ref{SpT_K_color}j).

When we compare the LTE model profiles with observations, the differences between H$\alpha$ and P$\delta$ are noticeable in both hot stars and, more significantly, in cool luminous stars. It is clear that the forbidden transition between the 1s and 2s levels plays a key role here.  The atomic hydrogen population in the 2s level is dynamically balanced through two channels. The first channel is the collisional, non-radiative transition.  The second is the two-step radiative transit through the np levels (1s$\leftrightarrow$np$\leftrightarrow$2s, n $> 2$).  In a low density environment such as the upper atmosphere of a very luminous giant star, collisional transition contributions to the 2s level decrease due to the low density. Our results seem to suggest that in the high temperature regime (A and early F type stars), this decrease in collisional transitions causes the 2s population to fall short of the LTE-predicted population. In the low temperature regime (G and later type stars), the low collision rate causes the 2s population to be greater than what is predicted by LTE.  For main sequence stars, the LTE population of 2s is proportional to $\exp(-E/kT)$, which decreases dramatically when the temperature drops. For luminous giants, our results suggest a shallower dependency of the 2s population to temperature than in main sequence stars.  In the atmospheres of luminous stars, as compared with main sequence stars, fewer hydrogen atoms are piled up in the $n=2$ levels in the high temperature regime while more hydrogen atoms are found in the same levels in the low temperature regime.  The difference disappears around the spectral type of F5.

\citet{prz04} point out that purely NLTE effects in the photosphere of Arcturus are not enough to make the model match the observation, and one must consider the NLTE contribution in an extended atmosphere model including the chromosphere. Generally, the inverted temperature structure of the chromosphere (increasing temperature outward) would lead to emission features. However, \citet{mih73} show that strong NLTE effects in the chromosphere can maintain the absorption in the H$\alpha$ region.

\section{Conclusion}	

We have observed a wide range of stars at the 10\,049 $\AA$ P$\delta$ line and a few stars at the 10\,938 $\AA$ P$\gamma$ line to make their line profiles available for comparison with model stellar atmospheres. Since the Paschen lines are less affected by the departures from LTE than the Balmer lines, they should be more convenient to use as a test of models. As expected, the difference in line profiles of Paschen lines as compared to H$\alpha$ is greatest for stars of high luminosity and low effective temperature.

\begin{acknowledgements}

We thank David Bohlender, Les Saddlemyer, Dmitry Monin, Marilyn Bell and other DAO staff for their assistance in making these observations successful. We are also very grateful for the financial support of the Kenilworth Fund of the  New York Community Trust.  This research has made use of the SIMBAD database, operated at CDS, Strasbourg, France. 

\end{acknowledgements}



\newpage
\clearpage

\longtab{1}{
\onecolumn
\begin{center}
\begin{longtable}{rlllccccp{0.52in}p{0.5in}}
\caption{\label{t1}Standard stars observed} \\
\hline \hline
HD&Sp. type&M$_{V}$&B-V&\multicolumn{3}{c}{Date of observation}&[Fe/H]&[Fe/H] reference&Online figure \\
\cline{5-7} \\
&&&&H$\alpha$&P$\delta$&P$\gamma$ \\
\hline
\endfirsthead

\multicolumn{10}{c}{{\tablename} \thetable{} -- \textit{Continued from previous page}} \\
\hline \hline
HD&Sp. type&M$_{V}$&B-V&\multicolumn{3}{c}{Date of observation}&[Fe/H]&[Fe/H] reference&Online figure \\
\cline{5-7} \\
&&&&H$\alpha$&P$\delta$&P$\gamma$ \\
\hline
\endhead

\hline \multicolumn{10}{c}{\textit{Continued on next page}} \\
\endfoot
\endlastfoot

37128&B0Iab&$-$7.21\tablefootmark{*}&$-$0.19&Feb. 5, 09&Feb. 5, 09&...&...&... &7\\
40111&B0.5II&$-$3.59&$-$0.09&Feb. 4, 09&Feb. 5, 09&...&...&... &8\\
91316&B1Iab&$-$7.31\tablefootmark{*}&$-$0.15&Dec. 5, 08&Feb. 3, 09&...&$-$0.89&1&9\\
2905&B1Iae&$-$6.47\tablefootmark{*}&0.09&Feb. 5, 09&Feb. 5, 09&...&...&... &10\\
24398&B1Ib&$-$3.91 &0.08&Feb. 5, 09&Feb. 5, 09&...&...&... &11\\
116658&B1III-IV&$-$3.53&$-$0.13&...&Jul. 1, 08&...&...&... &12\\
41117&B2Iae&$-$4.02&0.21&Feb. 5, 09&Feb. 5, 09&...&...&... &13\\
206165&B2Ib&...&0.22&Feb. 5, 09&...&...&$-$0.33&1&14\\
886&B2IV&$-$2.64&$-$0.23&Dec. 4, 08&Feb. 3, 09&...&$-$0.34&1&15\\
198478&B2.5Ia&$-$4.34&0.31&Dec. 4, 08&Jul. 2, 08&...&$-$0.23&1&16\\
35708&B2.5IV&$-$1.58&$-$0.14&Dec. 5, 08&Feb. 3, 09&...&...&... &17\\
14143&B3I&...&0.5&Feb. 4, 09&...&...&...&... &18\\
14134&B3Ia&...&0.44&Feb. 4, 09&...&...&...&... &19\\
11415&B3III&$-$2.19&$-$0.12&Dec. 4, 08&Feb. 3, 09&...&...&... &20\\
160762&B3IV&$-$1.97&$-$0.17&Dec. 4, 08&Jul. 2, 08&...&$-$0.4&2&21\\
36371&B4Ib&...&0.25&Dec. 5, 08&Feb. 3, 09&...&...&... &22\\
15497&B6Ia&...&0.64&Feb. 4, 09&...&...&...&...&23\\
155763&B6III&$-$1.88&$-$0.1&Jul. 4, 08&Jul. 2, 08&...&$-$0.95&3&24\\
3240&B7III&$-$1.03&$-$0.1&Dec. 4, 08&...&...&...&... &25\\
69686&B8&$-$1.10\tablefootmark{*}&$-$0.52&Feb. 4, 09&...&...&...&... &26\\
34085&B8Iab&$-$6.92&$-$0.03&Dec. 5, 08&Feb. 3, 09&...&...&... &27\\
14322&B8Ib&...&0.26&Feb. 4, 09&...&...&...&... &28\\
63975&B8II&$-$1.32&$-$0.11&Feb. 4, 09&Feb. 5, 09&...&...&... &29\\
21291&B9Ia&$-$4.56\tablefootmark{*}&0.34&Feb. 5, 09&Feb. 5, 09&...&...&... &30\\
212593&B9Iab&$-$4.58\tablefootmark{*}&0.07&Feb. 5, 09&...&...&...&... &31\\
46300&A0Ib&...&0&Feb. 5, 09&Feb. 5, 09&...&$-$0.3&4&32\\
87737&A0Ib&$-$4.45&$-$0.03&Dec. 5, 08&Feb. 3, 09&...&$-$0.05&4&33\\
47105&A0IV&$-$0.70&0.03&Dec. 5, 08&Feb. 3, 09&...&0.08&5&34\\
25642&A0IVn&$-$1.28&0.02&Dec. 5, 08&...&...&...&... &35\\
172167&A0V&0.66&0&Jul. 4, 08&Jul. 1, 08&Sep. 30, 09&$-$0.43&6&36\\
34787&A0Vne&0.18&$-$0.02&Dec. 5, 08&...&...&...&... &37\\
14433&A1Ia&...&0.58&Feb. 4, 09&...&...&...&... &38\\
214994&A1IV&0&$-$0.01&Dec. 4, 08&...&...&0.42&7&39\\
14489&A2Ia&$-$4.73\tablefootmark{*}&0.29&Feb. 4, 09&...&...&$-$0.26&4&40\\
197345&A2Iae&$-$6.89&0.09&Jul. 4, 08&Jul. 1, 08&...&0.15&8&41\\
50019&A3III&$-$0.18&0.1&Dec. 5, 08&Feb. 3, 09&...&...&... &42\\
17378&A5Ia&...&0.75&Dec. 26, 10\tablefootmark{\dagger}&Dec. 26, 10\tablefootmark{\dagger}&...&$-$0.07&9&43\\
2628&A7III&1.29&0.24&Dec. 4, 08&...&...&$-$0.24&6&44\\
31964&A8Iab&...&0.49&Feb. 4, 09&Feb. 3, 09&Sep. 30, 09&...&... &45\\
182835&F2Iab&$-$4.92\tablefootmark{*}&0.52&Dec. 4, 08&Jul. 1, 08&...&$-$0.03&9&46\\
20902&F5Iab&$-$4.06&0.47&Dec. 5, 08&Feb. 3, 09&Aug. 25, 10&$-$0.07&9&47\\
213306&F5Iab&$-$2.84&0.74&...&Jul. 2, 08&...&...&... &48\\
187929&F6Iab&...&0.74&...&Jul. 2, 08&...&0.01&10&49\\
33564&F6V&3.59&0.45&Dec. 5, 08 &...&...&0.21&11&50\\
215648&F7V&3.25&0.49&Dec. 4, 08&...&...&$-$0.35&12&51\\
18391&G0Ia&...&1.87&Dec. 26, 10\tablefootmark{\dagger}&Dec. 26, 10\tablefootmark{\dagger}&...&...&... &52\\
204867&G0Ib&$-$3.04&0.83&Dec. 4, 08&Jul. 1, 08&...&0.1&9&53\\
209750&G2Ib&$-$2.92&0.96&Dec. 4, 08&Jul. 2, 08&...&0.03&9&54\\
188727&G5Ibv&...&0.88&Dec. 4, 08&Jul. 2, 08&...&...&... &55\\
71369&G5III&$-$0.19&0.83&Dec. 5, 08&Feb. 3, 09&...&$-$0.16&13&56\\
77912&G8Ib-II&$-$2.27&1.02&Dec. 5, 08&Feb. 3, 09&...&0.38&14&57\\
62345&G8IIIa&0.54&0.93&Dec. 5, 08&Feb. 3, 09&...&$-$0.02&13&58\\
6833&G9III&0.31&1.14&Dec. 26, 10\tablefootmark{\dagger}&Dec. 26, 10\tablefootmark{\dagger}&...&$-$0.89&15&59\\
124897&K1.5III&$-$0.15&1.23&Jul. 4, 08&Jul. 2, 08&...&$-$0.52&16&60\\
206778&K2Ib&$-$4.08&1.56&Dec. 4, 08&Jul. 1, 08&Sep. 30, 09&$-$0.18&13&61\\
85503&K2III&1.15&1.22&...&Jul. 2, 08&...&0.29&13&62\\
156283&K3Iab&$-$2&1.46&Jul. 4, 08&Jul. 2, 08&...&0.01&13&63\\
17506&K3Ib&$-$3.26&1.68&Feb. 4, 09&...&...&0.09&17&64\\
44537&K5Iab&...&1.96&Feb. 5, 09&Feb. 5, 09&...&0.08&18&65\\
29139&K5III&$-$0.55&1.48&...&...&Sep. 30, 09&$-$0.15&16&66\\
147767&K5III&$-$0.78&1.57&...&Jul. 1, 08&...&$-$0.16&19&67\\
42543&M0Iab&...&2.22&Feb. 5, 09&Feb. 5, 09&...&$-$0.42&20&68\\
6860&M0III&$-$1.74&1.58&Dec. 4, 08&Jul. 2, 08&Sep. 30, 09&$-$0.04&21&69\\
14330&M1sd&...&2.07&Dec. 26, 10\tablefootmark{\dagger}&Dec. 26, 10\tablefootmark{\dagger}&...&...&... &70\\
206936&M2Ia&...&2.26&Aug. 25, 10&...&Aug. 25, 10&...&... &71\\
36389&M2Iab&$-$4.36&2.07&Feb. 5, 09&Feb. 5, 09&Sep. 30, 09&0.11&20&72\\
39801&M2Iab&$-$5.42&1.85&Dec. 5, 08&Feb. 3, 09&Sep. 30, 09&0&21&1\\\\
13136&M2Iab-b&...&2.29&Dec. 26, 10\tablefootmark{\dagger}&Dec. 26, 10\tablefootmark{\dagger}&...&...&... &73\\
147749&M2III&$-$1.19&1.64&Jul. 4, 08&...&...&...&... &74\\
217906&M2.5II-III&$-$1.41&1.67&Aug. 25, 10&Jul. 2, 08&Sep. 30, 09&$-$0.11&22&75\\
14469&M3Iab&...&2.12&...&Jul. 2, 08&...&...&... &76\\
42995&M3III&$-$2.06&1.59&...&...&Sep. 30, 09&0.04&16&77\\
44478&M3III&$-$1.34&1.62&...&...&Sep. 30, 09&0&16&78\\
172380&M4Iab&$-$2.74\tablefootmark{*}&1.44&Aug. 25, 10&...&Aug. 25, 10&...&... &79\\
197812&M5Iab&...&1.42&Aug. 25, 10&...&Aug. 25, 10&...&... &80\\
156014&M5Ib-II&$-$2.3&1.45&Jul. 4, 08&Jul. 1, 08&...&...&... &81\\
\hline
\end{longtable}
\tablefoot{
\tablefoottext{*}{The parallax for this star lies between 3 and 5 times its probable error.}
\tablefoottext{\dagger}{Data obtained at the Apache Point Observatory.}
}
\tablebib{
(1)~\citet{gies92}; (2) \citet{peters70}; (3) \citet{smith93}; (4) \citet{venn95}; (5) \citet{hill95}; (6) \citet{erspamer03}; (7) \citet{hui00}; (8) \citet{albayrak00}; (9) \citet{lyubimkov10}; (10) \citet{luck81}; (11) \citet{gonzalez10}; (12) \citet{fuhrmann08}; (13) \citet{hekker07}; (14) \citet{fernandez90}; (15) \citet{mishenina01}; (16) \citet{melendez08}; (17) \citet{mallik98}; (18) \citet{bakos71}; (19) \citet{mcwilliam90};  (20) \citet{luck80}; (21) \citet{carr00}; (22) \citet{smith85}.
}
\end{center}
}

\begin{table*}[h]
\centering
\caption{\label{t2}Metal-poor stars observed\tablefootmark{\dagger}} 
\begin{tabular}{r l l l l c c c p{0.5in} p{0.5in}}
\hline \hline
HD&Name&Sp. type&M$_{V}$&B-V&\multicolumn{2}{c}{Date of observation}&[Fe/H]&[Fe/H] reference&Online figure \\
\cline{6-7} \\
&&&&&H$\alpha$&P$\delta$ \\
\hline
2796&BD$-$17 70&Fp&...&0.67&Nov. 21, 10&Nov. 21, 10&$-$2.21&1&82\\
140283&BD$-$10 4149&F3sd&3.47&0.5&May 9, 09&May 9, 09&$-$2.38&2&83\\
6755&BD+60 170&F8V&2.14&0.67&Aug. 21, 10&Aug. 21, 10&$-$1.47&1&84\\
221170&BD+29 4940&G2IV&0.18\tablefootmark{*}&1.02&Aug. 21, 10&Aug. 21, 10&$-$2.05&1&85\\
216143&BD$-$07 5873&G5&...&0.94&Aug. 21, 10&Aug. 21, 10&$-$2.11&1&86\\
2665&BD+56 70&G5IIIp&0.36\tablefootmark{*}&0.71&Dec. 26, 10&Dec. 26, 10&$-$1.97&2&87\\
...&BD+44 493&G5IV&2.62\tablefootmark{*}&0.52&Dec. 26, 10&Dec. 26, 10&$-$3.7&3&88\\
218732&BD$-$14 6415&G7Ib&...&1.66&Nov. 21, 10&Nov. 21, 10&$-$1.64&4&89\\
187111&BD$-$12 5540&G8p&...&1.17&Jul. 23, 10&Jul. 23, 10&$-$1.74&1&90\\
122563&BD+10 2617&K0IIp&$-$0.55&0.9&May 5, 09&May 5, 09&$-$2.71&5&91\\
103095&BD+38 2285&K0Vp&6.77&0.75&Jun. 8, 09&Jun. 8, 09&$-$1.35&2&92\\
4306&BD$-$10 155&KIIp&2.65\tablefootmark{*}&0.63&Nov. 21, 10&Nov. 21, 10&$-$2.52&1&93\\
25329&BD+34 796&K1V&7.33&0.87&Nov. 21, 10&Nov. 21, 10&$-$1.84&6&94\\
165195&BD+03 3579&K3p&...&1.24&Jul. 23, 10&Jul. 23, 10&$-$2.03&1&95\\
232078&BD+16 3924&K3IIp&...&2.03&Jul. 23, 10&Jul. 23, 10&$-$1.6&3 \& 4&96\\

\hline
\end{tabular}
\tablefoot{
\tablefoottext{*}{The parallax for this star lies between 3 and 5 times its probable error.}
\tablefoottext{\dagger}{Data obtained at the Apache Point Observatory.}
}
\tablebib{
(1)~\citet{mishenina01}; (2) \citet{bergemann08}; (3) \citet{barzdis10}; (4) \citet{ito09}; (5) \citet{burris00}; (6) \citet{bergemann08}.
}
\end{table*}

\begin{table*}
\begin{center}
\caption{\label{t3}Atomic data for hydrogen lines}
\begin{tabular}{l l p{0.5in} p{0.5in} l}
\hline \hline
Line&Wavelength ($\AA$)&Lower energy level&Upper energy level&log\emph{gf} \\
\hline
H$\alpha$&6\,562.80&$2^{2}$$p^{0}$&$3^{2}$d&+0.71 \\
P$\delta$&10\,049.4&$3^{2}$d&$7^{2}p^{0}$&$-$0.30 \\
P$\gamma$&10\,938.1&$3^{2}$d&$6^{2}p^{0}$&+0.00\\
\hline
\end{tabular}
\end{center}
\end{table*}

\begin{table*}
\centering
\begin{center}
\caption{\label{t4}Model hydrogen line profile parameters} 
\begin{tabular}{r c l c c c c}
\hline \hline
Figure&Star name&HD&$T_{\rm eff}$&log{\emph g}&$V_t$(km/s)&[Fe/H] \\
\hline
\ref{early_LumI_color}b \& \ref{early_LumI_color}h&13 Mon&46300&9\,700&2.0&4.0&$-$0.1 \\
\ref{rev_HD172167_color} \& \ref{pgamma_color}a&Vega&172167&9\,540&4.0&1.2&$-$0.5 \\
41\tablefootmark{*}&Deneb&197345&9\,100&1.5&6.0&0.0 \\
\ref{early_LumI_color}c \& \ref{early_LumI_color}i&...&182835&6\,750&1.0&5.0&0.0 \\
\ref{late_LumI_color}c \& \ref{late_LumI_color}i&$\alpha$Per&20902&6\,350&1.9&5.3&0.0 \\
\ref{early_LumI_color}d \& \ref{early_LumI_color}j&...&209750&5\,250&1.5&3.8&0.0 \\
\ref{SpT_K_color}e \& \ref{SpT_K_color}k&Groombridge 1830&103095&4\,950&4.5&0.7&$-$1.3 \\ 
58\tablefootmark{*}&...&62345&4\,800&2.9&3.8&$-$0.3 \\
95\tablefootmark{*}&...&165195&4\,430&1.2&1.9&$-$2.1 \\
\ref{SpT_K_color}b \& \ref{SpT_K_color}h&...&122563&4\,425&0.6&2.1&$-$2.6 \\
\ref{SpT_K_color}d \& \ref{SpT_K_color}j&Arcturus&124897&4\,250&1.9&1.4&$-$0.5 \\
\ref{SpT_K_color}c \& \ref{SpT_K_color}i&...&232078&4\,000&0.3&2.6&$-$1.6 \\
\ref{rev_HD39801_color}&$\alpha$Ori&39801&3\,540&0.0&2.3&0.0 \\
\hline
\end{tabular}
\tablefoot{
\tablefoottext{*}{Figure is available electronically only.}
}
\end{center}
\end{table*}
\clearpage
\newpage

\onlfig{1}{
\begin{figure} [H]
\resizebox{\hsize}{!}{\includegraphics{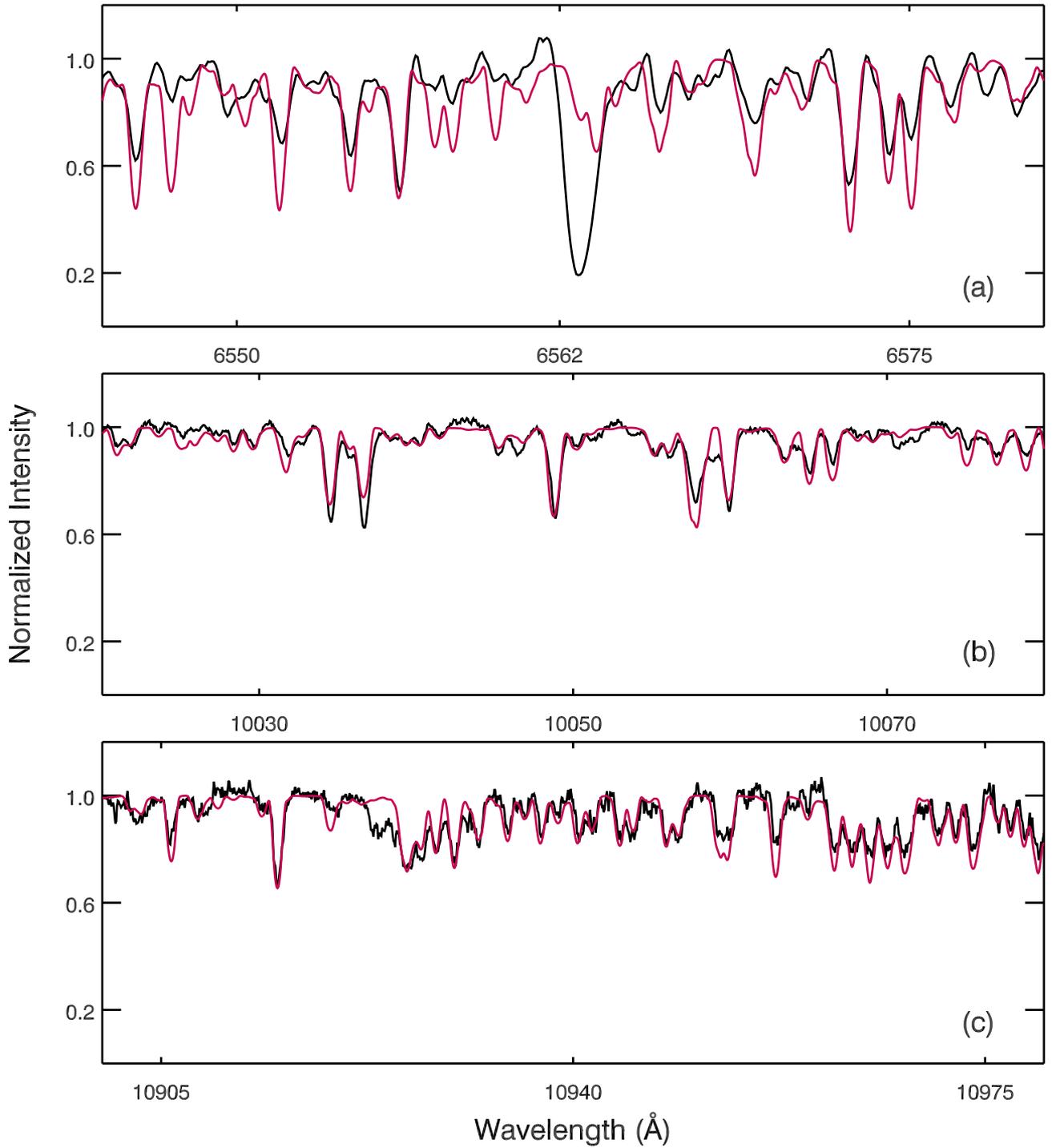}}
\caption{Profiles for H$\alpha$, P$\delta$, and P$\gamma$ (Fig. (a), (b), and (c) respectively) in HD39801 ($\alpha$Ori) M2 Iab. The black line is the observed spectrum and the red line is the model spectrum.}
\label{rev_HD39801_color}
\end{figure}
}

\onecolumn

\onlfig{2}{
\begin{figure*}
\centering
\resizebox{\hsize}{!}{\includegraphics{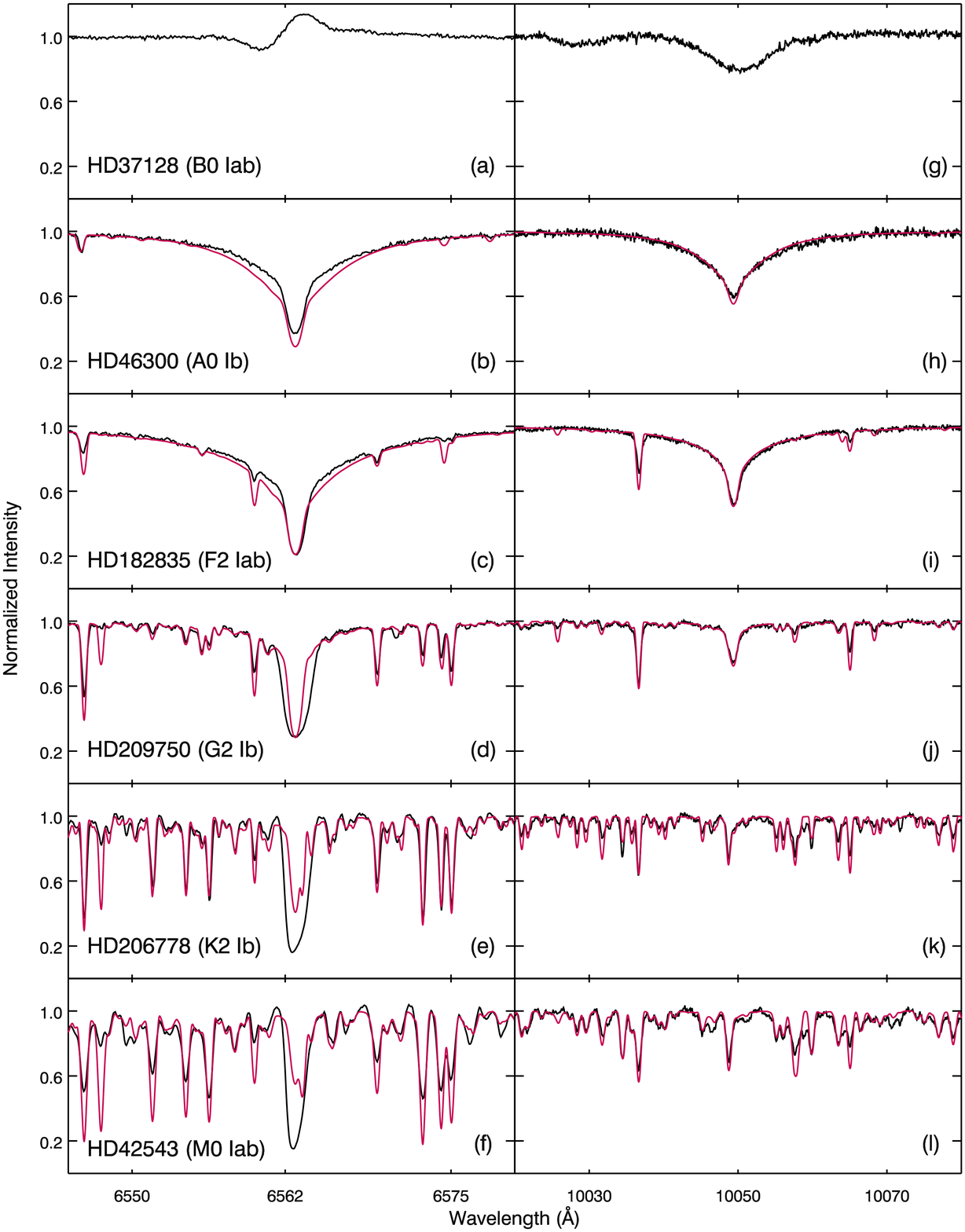}}
\caption{Profiles of H$\alpha$ (Fig. (a)--(f)) and P$\delta$ (Fig. (g)--(l)) for selected stars of high luminosity.}
\label{early_LumI_color}
\end{figure*}
}

\onlfig{3}{
\begin{figure*}
\resizebox{\hsize}{!}{\includegraphics{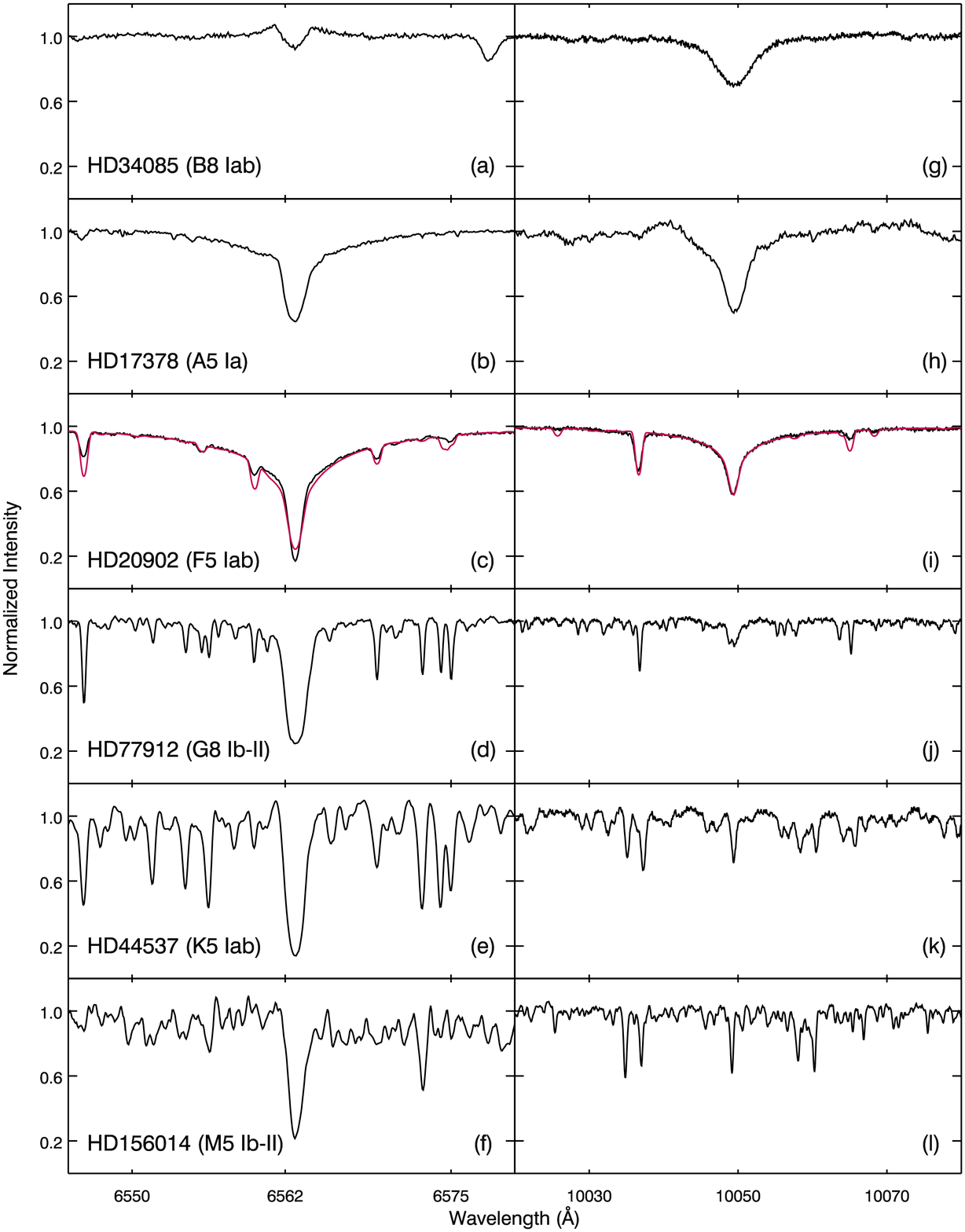}}
\caption{Profiles of H$\alpha$ (Fig. (a)--(f)) and P$\delta$ (Fig. (g)--(l)) for selected stars of high luminosity.}
\label{late_LumI_color}
\end{figure*}
}

\onlfig{4}{
\begin{figure*}
\resizebox{\hsize}{!}{\includegraphics{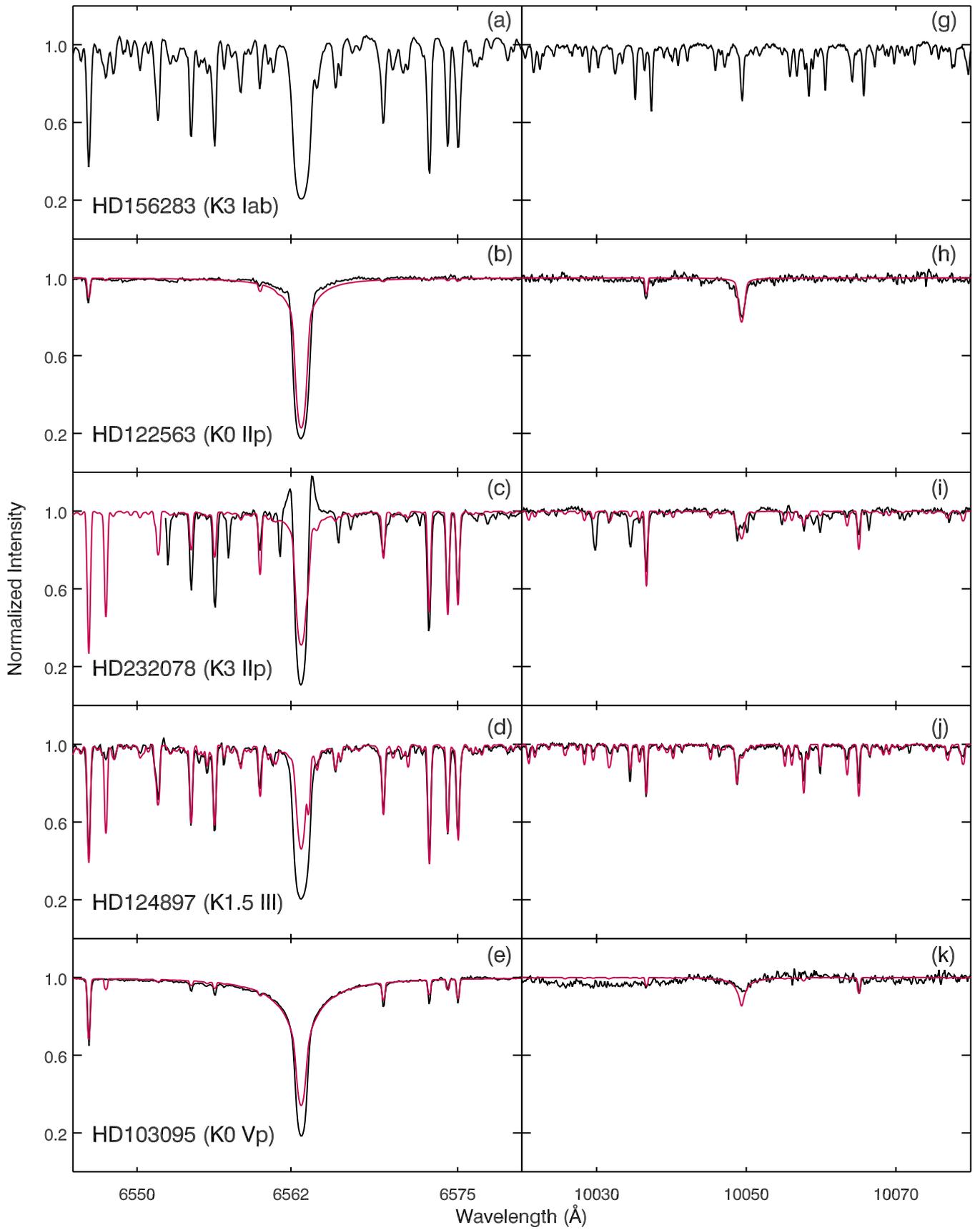}}
\caption{Profiles of H$\alpha$ (Fig. (a)--(e)) and P$\delta$ (Fig. (g)--(k)) for selected spectral type K stars.}
\label{SpT_K_color}
\end{figure*}
}

\onlfig{5}{
\twocolumn
\begin{figure}[H]
\resizebox{\hsize}{!}{\includegraphics{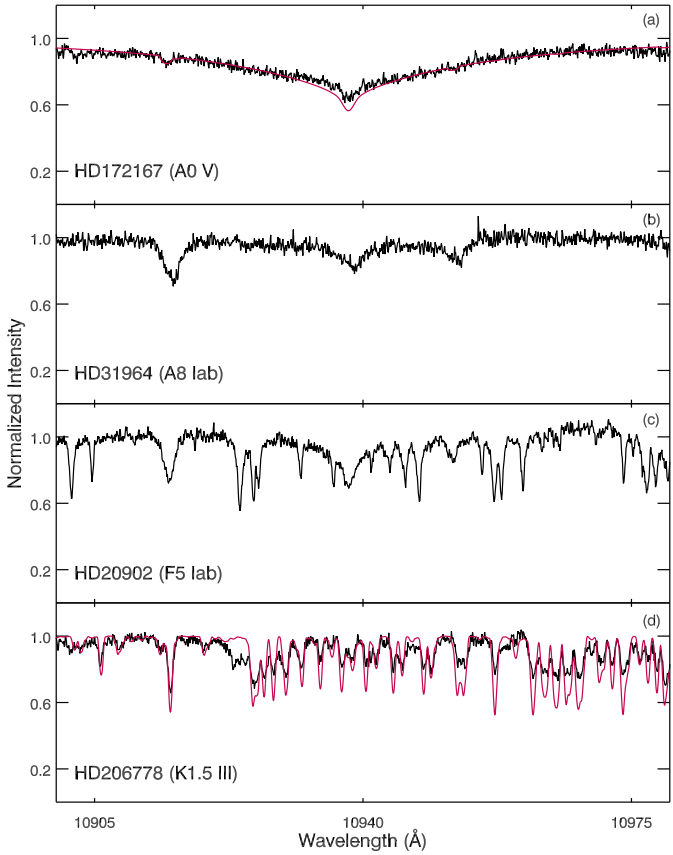}}
\caption{Profiles of P$\gamma$ for select stars.}
\label{pgamma_color}
\end{figure}
}

\onlfig{6}{
\begin{figure}[H]
\resizebox{\hsize}{!}{\includegraphics{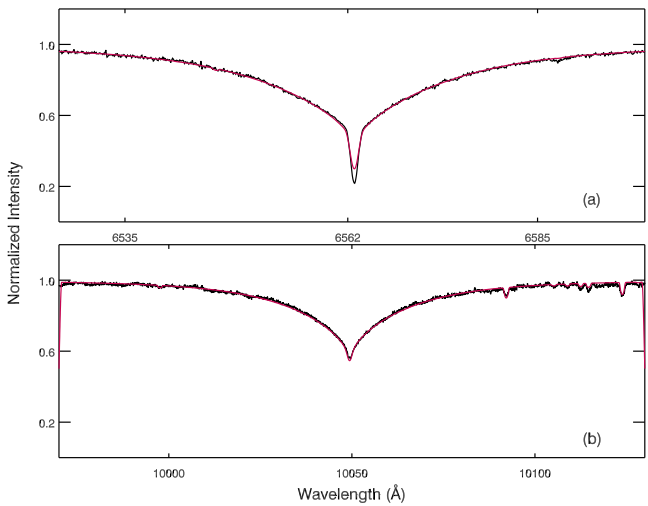}}
\caption{HD172167 (Vega) A0 V: Except for a very small deviation at the center of the H$\alpha$ profile, this fundamental standard for modelling stellar atmospheres shows a very good fit between the modelled and the observed profiles for H$\alpha$ and P$\delta$.}
\label{rev_HD172167_color}
\end{figure}
}



\end{document}